# Rural Wireless Mesh Network: A Design Methodology

**Jean Louis Ebongue Kedieng Fendji[1], Jean Michel Nlong[2]**

[1]University Institute of Technology (UIT), The University of Ngaoundéré, Ngaoundéré, Cameroon
[2]Faculty of Science, The University of Ngaoundéré, Ngaoundéré, Cameroon
Email: lfendji@yahoo.fr, jmnlong@yahoo.fr

## Abstract

**Wireless Mesh Network is presented as an appealing solution for bridging the digital divide between developed and under-developed regions. But the planning and deployment of these networks are not just a technical matter, since the success depends on many other factors tied to the related region. Although we observe some deployments, to ensure usefulness and sustainability, there is still a need of concrete design process model and proper network planning approach for rural regions, especially in Sub-Saharan Africa. This paper presents a design methodology to provide network connectivity from a landline node in a rural region at very low cost. We propose a methodology composed of ten steps, starting by a deep analysis of the region in order to identify relevant constraints and useful applications to sustain local activities and communication. Approach for planning the physical architecture of the network is based on an indoor-outdoor deployment for reducing the overall cost of the network.**

## 1. Introduction

The development of a region depends on not only the circulation of persons but also on the circulation of information. In a world that claims to be a global village, good information traffic is one of the biggest concerns. Because it enables development of business models, prevention of disasters or diseases, implementation of new learning process and so on. But many rural regions and sometime urban and semi-urban regions in developing countries are still suffering from the lack of connectivity. That hinders the development, resulting in degradation of both social communication and business advances in those regions and the digital divide becomes more and more apparent. As the map of the Internet World Stats can show, African Internet users represent only 8.6% of the world Internet users in 2013 [1].

Connecting rural regions in developing countries is a hard task because of many reasons. The first is their location with sometimes the lack of proper roads and hostile environment that makes them difficult to reach. Another reason is the lack, the deficiency or the instability of infrastructure, mainly power infrastructure; that seriously hinders the deployment of technology since an autonomous way of powering devices and network equipment is required. The cost of the bandwidth is still expensive especially in Africa, though cellular or satellite coverage is available in these rural regions. The lack of local capacity in network administration and maintenance in these regions is another important inhibiting factor. Beside these constraints [2] reveals other ones like language and cultural barriers, transportation issues, tampering and theft (in some areas). Because network service providers are driven by profit, the low-density population with low-income in these regions cannot insure a return of investment. Moreover, in some countries, the lack of awareness about the potential of wireless networks to support the development of rural regions, leads the government to do not define a clear policy to solve this problem. In contrary, law and regulations, like licensed spectrum policies, constitute a real barrier for providing some solutions.

However, despite the impediments observed in rural regions, there are several reasons to consider these regions, especially in Africa. The first reason is that those regions host the majority of the population. According to [3], in the second half of 2012, the percentage of African population living in those regions is estimated to 60.1 percent. This represents a great market; and to take advantage, new business models should be developed.

The second reason is the rural exodus. The emerging policy of some African developing countries essentially relies on the agriculture that is mostly done in rural region. Therefore, rural exodus could seriously hinder their development. A report from U. N. Habitat [4] states that 14 million people in Sub-Saharan Africa migrate from rural to urban regions every year. Considering this migration, the report predicts that more Africans will live in urban than in rural regions by 2030. This constitutes a real danger for the economy of these countries since it heavily relies on agriculture.

Although basic human needs are not completely satisfied in those regions, reaching information can provide access to education, to health information, help preventing disaster and so on. Some examples show how simple access to information can transform impoverished regions: Fishing industry in India, sunflower farming in Zambia. Even some social problems such as gender inequality can be overcome [5].

## 2. Rural Cameroon and Multipurpose Community Telecentres

Cameroon is considered as Africa in miniature. Indeed, the climate diversity, the different cultures and reliefs of its rural regions expose different characteristics. With 1.20% of Internet's contribution to the overall Gross Domestic Product GDP [6], Cameroon is the eighth country in Africa where Internet could bring more positive changes in daily life of the population and foster the national development. **Table 1** provides an insight of the telecommunication situation in Cameroon according to ITU [7].

In Cameroon we count 330 townships with 305 rural ones (more than 92%). But rural population represents only 41.60% in 2010, compared to 47.46% in 2002 [8]. This diminution of the rural population is mainly caused by the rural exodus, as it is the case in Africa in general. This situation is a hindrance to Cameroon which intends to become an emerging country at the horizon of 2035. This is because its emerging policy relies largely on agriculture which is mainly done in these rural regions. Beside this objective to become an emerging country, one of the greatest projects from the Ministry of Telecommunication in Cameroon is the creation of Multipurpose Community Telecentres (MCTs). These are access points designed to provide ICT services, as well as postal and financial services to a community, at low prices. They are designed to be deployed in rural and suburban regions to fight against the digital divide between these regions which are generally neglected by private network service providers and urban regions which are more attractive in terms of return on investment. **Figure 1** shows the different services provided by a MCT.

**Table 1.** Telecommunication indicators in Cameroon, Africa, World.

| Indicators | Regions | | |
|---|---|---|---|
| | Cameroon | Africa | World |
| Fixed-telephone subscriptions (2013) | 3.59% | 1.30% | 16.20% |
| Mobile-cellular telephone subscriptions (2013) | 70.39% | 65.90% | 93.10% |
| Fixed (wired)-broadband subscriptions (2013) | 0.08% | 0.30% | 9.40% |
| Percentage of individuals using the Internet (2013) | 6.40% | 16.80% | 37.90% |
| Internet contribution to GDP (2012) | 1.20% | 1.10% | 3.70% |

In 2012, we counted around 50 operational MCTs; 117 MCTs in launching phase; 28 in construction; four operational community radio stations and 10 community radio stations in construction. In the Adamawa region in Cameroon, we count 13 MCTs among which 06 are operational and 03 are functional. **Table 2** provides some information about the localities where three MCTs are deployed in the Adamawa region.

MCTs are frequently closed for diverse reasons; mainly depending on the availability of the manager or the availability of resources. The low number of connected machines cannot serve all the clients at the same time. This makes the clients lining up. An important reason of the failure of MCTs is that information is not relevant for local population; the needs of this population are not meet.

One of the objectives of these MCTs is to cover a radius of about 35 km. Achieving this perspective requires a careful design and deployment. Among questions raised by this perspective and the actual state of MCTs, we can list: How can this signal be spread from a landline node at very low cost in a rural community? How this network can be useful and sustainable for the local community? And how can the information be relevant for the local population? The main objective of this paper is to present a design methodology for providing a sustainable local network from a landline node in a rural region at very low cost.

## 3. Related Work on Rural Network Planning and Deployment

Wireless Mesh Networks (WMN) emerges as a solution to realize the dream to connect rural regions to the rest of the world. Indeed, WMNs can easily, effectively and wirelessly connect entire cities or villages both locally and to Internet, using inexpensive existing technology.

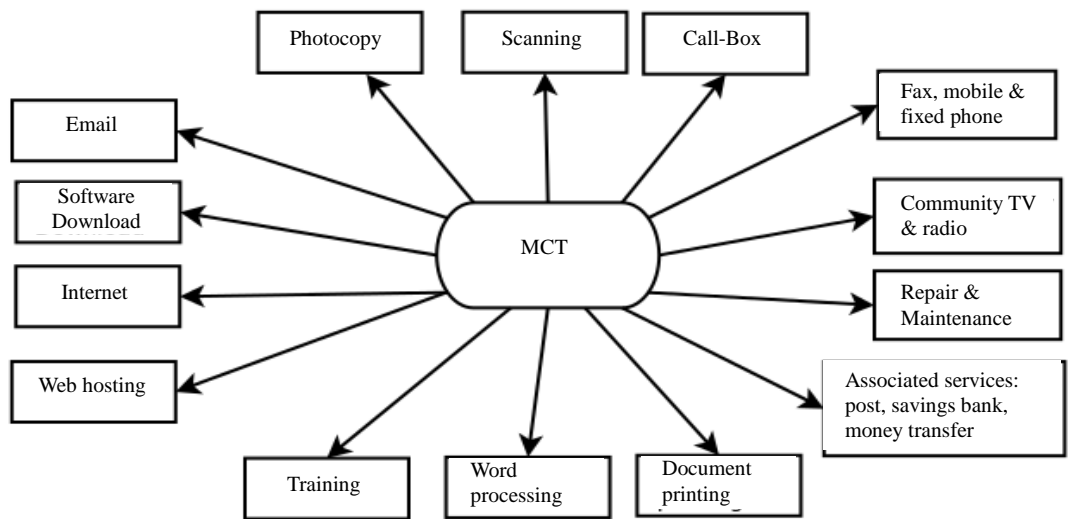

**Figure 1.** Multipurpose community telecentre services.

**Table 2.** Some telecentres in Adamawa region in Cameroon.

| Locality | MCT Powering | Population | Area in km² | Main Activities |
|---|---|---|---|---|
| Mbé | Generator | 43,000 | 3000 | Agriculture |
| Ngaoui | Solar Panel | 15,000 | 2307 | Trading and Farming |
| Tibati | Grid | 46,563 | 8000 | Fishing and Farming |

The first step in building a WMN is the network planning. But this step received few attentions over the last few years; despite the plethora of studies that has been carried out on WMNs. Earlier WMN deployments did not use systematic approach in the planning. Consequently, these deployments have experienced a number of shortcomings tied to the connectivity problems (dead spots, lack of coverage) and performance problems (high latency and low throughput). Moreover, the scalability of the network could not be guaranteed and the overall cost of the system that is the main concern especially in rural regions, did not received attention and therefore could not be estimated.

Recently, there has been an emphasis on developing wireless networks for rural regions. The focus was to carefully design long link wireless networks called Wifi Long Distance (WiLD) [9] [10]. A typical deployment consists in linking a nearby city to a so called kiosk located in a remote village.

Sen et al. [9] provide the first relevant work on this field in rural areas by formulating the problem in terms of variables, constraints and the optimization criterion. The optimisation problem is to minimise the overall cost of the network affected by the multi-hop network topology and the antenna tower heights under three constraints: throughput, interference, and power.

In the same direction, Dutta et al. [10] provide another geometric formulation of the topology construction problem: given the location of the nodes, direct links between which nodes should be established such that, 1) all nodes are connected with certain robustness in the network, and 2) the cost of constructing towers to establish all the selected links is minimised. But they do not provide any algorithm to solve it.

Panigrahi et al. [11] provide amelioration of both [9] [10]. For [11], they provide an algorithm with the worst-case logarithmic bound for prior heuristic. They provide an approach that can handle any number of obstructions between two nodes, while the prior heuristic considers only one obstruction. Finally, they provide algorithms that can handle a much more general cost function compared to the piece-wise linear cost function for the antenna towers associated to the prior heuristic. For [10], Panigrahi et al. provide the first algorithms for solving the topology construction problem with provable guarantees on the approximation factor.

Although this prior kiosk-based model results are satisfactory, the increase of user connectivity demand and the possibility for developing new business models, in order to bridge the digital divide, require more than just a kiosk model. Therefore, the problem is to spread this signal in a rural region after providing the connectivity to a point of interest. Some works has been already done in the direction of rural local access network deployment [12]-[15].

FRACTEL [13] partially tackles this problem by considering both long-distance network (LDN) and local access networks (LACNs). In this work, Chebrolu and Raman provide novel approaches to TDMA scheduling and channel allocation, but they do not provide a real approach for LACN planning.

Considering only local access network, Naidoo and Sewsunker [15] evaluate the applicability of star mesh network solutions to cover a rural region of South Africa, using the 802.11 g standard. The main concern is to maximise the coverage, given a user node throughput (100 Kbps). But they also do not provide a concrete local planning approach.

[12] proposes and analyses the performance of an indoor to indoor community based rural WMN using 802.11 equipment. Since in rural areas we usually observe a low density population constituted in sparse group of houses, to connect a community from a landline node, using indoor to indoor deployment, seems too difficult. Because indoor to indoor deployment could result in separated sub networks, due to often long distances between groups of house.

Another work is [16] where the authors study efficient mesh router placement in WMN. Their mesh router placement problem is the determination of a minimum set of positions among the candidate positions in such a way that the mesh routers situated in these positions cover the given region, maintain the full connectivity toward the internet gateway and meet the traffic demand.

To obtain an efficient and sustainable network, the planning process should take into consideration more than just the technical aspect. Szabo et al. [17] suggest a design methodology going from the identification of applications and services to the cost calculation of the final network. But this design methodology is more adapted to city than to rural regions, since it does not consider, in a particular manner, sparsely populated regions with low income revenue. Mendez-Range and Lozano-Garzon [18] provide a design methodology for rural areas of developing countries, but this one is specifically designed for e-health services. To support local development, a rural multi-purpose network should be planned. It should support local activities, education, health services while considering the constraints of the region.

## 4. Design Methodology

### 4.1. Approach

The aim of our work is to spread the signal in a village from a landline node, by avoiding towers for reasons of cost minimization. Planning a WMN especially in rural regions is not just a technological or technical matter, because its success and sustainability depends on its integration in the daily life of the end users. This integration can be achieved only if the planned network finds a trade-off between what users need and what they could afford.

Our approach is based on two observations:

1) Using only outdoor nodes for the network is really costly (extra cost of mast and energy supply);

2) Using only indoor nodes is not realistic, since we usually observed sparse group of houses that could result in separated sub networks.

Therefore one idea could be to plan a network composed of both indoor and outdoor nodes.

Although avoiding antenna towers reduces significantly the cost of the network deployment, the outdoor deployment is still costly, because it requires usage of mast, independent way of powering (solar panel or battery) that can be reduced or avoided by using rooftop and grid power in indoor deployment. Moreover, the security aspect provided by indoor deployment is not trivial.

We mainly differ from prior works by defining a mixed approach (indoor-outdoor). We therefore distinguish two node types: indoor and outdoor. In the respect of coverage and throughput, wherever possible, an indoor node is preferred to an outdoor one for cost minimization reasons.

Because we are avoiding towers, cost functions defined in prior works are no longer adapted. There is a need to define a new cost function.

Finally, always in the scope of minimizing the overall cost of network deployment, the total coverage planning of a given region may be reduced, since we usually observe large spaces between groups of houses. Therefore, for a given a community, areas of interest could be defined. For sustainability reasons, geographical dispersion and population growth rate and income should be taken into consideration.

### 4.2. Design Steps

Our design process model is illustrated in **Figure 2**—partially similar to that in [17].

**Step 1: Analysing Regions**

This question is not only geographical. We have to also deal with socio-economic aspect. First, we select two zones based on the localisation and ethnic group (culture). As [2] suggests, social aspect (especially cultural) and natural environment influence the success of the implementation of a wireless network in rural regions.

After selecting the different areas, we do a survey based on the following criteria:

- Geo-economic considerations: Density of population, main activities, population incomes and growth rate (for sustainability), topology and surface, climate. Population growth can easily bring more changes in the environment and also in term of network requirement.
- Infrastructural considerations: Roads, energy supply, building, markets, and Internet access points;
- Technological considerations: Computer, mobile device (cell phone, tablet PC…), network operator, radio and TV spectrum availability (white spectrum availability);
- Governmental considerations: Spectrum regulation, other laws and regulations, government interest (services and future projects);
- Cultural considerations: Cultural barriers to technology penetration;
- Skill considerations: Local competence, ease of troubleshooting, people motivation.

For the survey, questionnaires are used and Interviews are performed. This may require an interpreter. Data are collected and synthesized in order to provide relevant information for the rest of the design process.

**Step 2: Identifying Coverage Requirements**

In rural region, there is no need to cover a whole region. So, to minimize the cost, only areas of interest should be covered. At this step, we circumscribe Areas of Interest (AI), where the signal must be spread in the region and we decompose the considered region into elementary areas. So, each elementary area can be characterised by:

- The interest: It can be of interest (where the signal must be spread like a market, a school, a hospital…) or not that means covering it is optional.

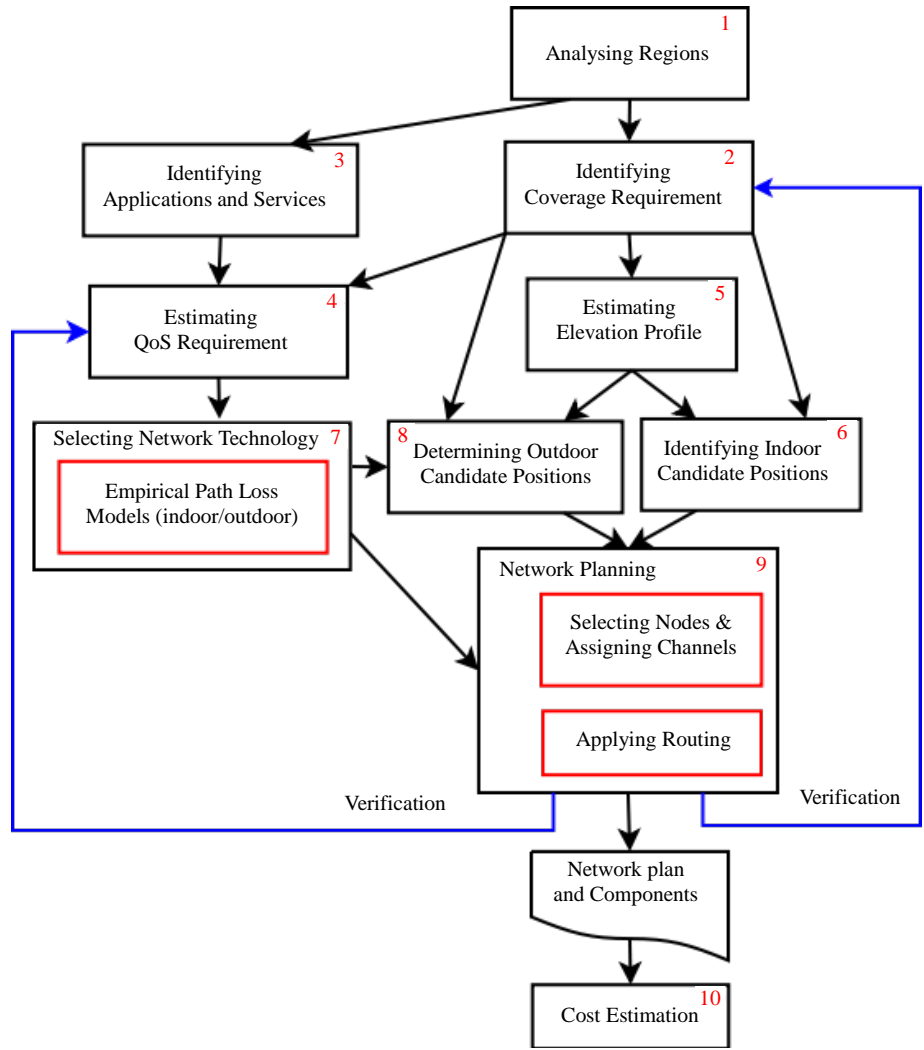

**Figure 2.** Design process model.

- The placement: It can be a place that can host a node or not (like road, lake …).
- The type of node: It can host an indoor or an outdoor node.

These characteristics are later used for selecting among candidate positions, the ones where the nodes will be installed.

For sustainability reasons, for each elementary area of interest ei, we define two factors affecting ei: Geographical dispersion factor di belonging to D and population growth factor gi belonging to G.

In this model, the universe is considered as a discrete set of predefined positions.

**Step 3: Identifying Applications and Services**

Here, we identify relevant applications and services the network should provide and their requirements (especially the throughput). We try to estimate the need of one user in term of throughput. An extension of this step is to provide a framework for the design and deployment process of these applications and services on the network. For example, since the bandwidth is usually low in these regions, using caches to avoid huge amount of external data traffic on network would be an interesting option.

**Step 4: Estimating QoS requirement**

A trivial way to estimate the QoS requirement, considered here as the throughput only, is to multiply the application and service requirement for one user by the number of simultaneous users. A more realistic approach is to estimate the budget of every link and to determine the budget at the gateway. **Table 3** gives some requirements according to type of application that will be deployed on the network [19].

**Table 3.** Application requirements.

| Type of Application | Requirements | Information |
|---|---|---|
| Text Messaging | <1 Kbits/s | Non-connected and tolerate high latency |
| Email | 1 to 100 Kbits/s | Non-connected and tolerate latency |
| Web Browsing | 50 to 100 Kbits/s | Not connected and tolerate latency |
| Audio Streaming | 96 to 160 Kbits/s | Connected constant amount of relatively large bandwidth |
| Voice over IP | 24 to 100 Kbits/s | Connected constant amount of bandwidth |
| Video Streaming | 64 to 200 Kbits/s | Connected high throughput and low latency |
| Peer to Peer | 0 to max throughput | Tolerate any amount of latency, but tend to use up all available throughput |

**Step 5: Estimating Elevation Profile**

After circumscribing areas of interest, we need to estimate the elevation profile. This is required for outdoor nodes since they will use a mast as support. The elevation profile will provide the height of outdoor nodes. Because there is no great change in term of price, we will assume that all outdoor nodes have the same mast height. The Fresnel zone must be considered when calculating the mast height. We want to check here that 10 meters is enough for a mast. A more realistic approach is to determine for each EAI ei the associated profile elevation.

**Step 6: Identifying Indoor Candidate Positions (ICP)**

ICP are chosen among houses or buildings which provide a grid power and help to avoid a high mast. These points reduce the price of deploying an indoor node, when compare to an outdoor node. Normally, all building that can be used as a support and provide grid power is seen as a candidate. However, this trivial condition can be reconsidered in order to obtain only relevant ICPs. Relevant characteristics of an ICP are the elevation profile that can help avoiding long mast and the reliability of grid power that can avoid long shortages.

**Step 7: Selecting Network Technology (Empirical Path Loss Models)**

The more appropriated technology is 802.11 (WiFi) because of his lower price when compared to 802.16 (WiMAX) which in turn provides a greater coverage but requires more complex infrastructure. 802.11 standards have many standards; however we focus on only two: 802.11g and 802.11n. 802.11g has a theoretical transmission range of 150 meters with a throughput of 54 Mbps while 802.11n has a theoretical transmission range of 250 meters with a throughput of 248 Mbps. We study the empirical path loss models in rural environment by defining three scenarios: Free space (without obstacles), Raised space (foliage), Built space (Houses). The aim here is to provide a precise empirical path loss model tied to rural areas at 2.4 GHz. Results are compared to existing method of prediction and the best is optimized.

**Step 8: Identifying Outdoor Candidate Positions (OCP)**

OCPs depend on the results of the empirical path loss models of Step 7, because they are determined using these results and coverage requirements from Step 2. From the gateway, we will determine all the possible OCPs according to the transmission range.

**Step 9: Network Planning**

It is the most important step in the design process. It is composed of two main phases: Selecting nodes and assigning channel, and applying routing protocol.

Selecting nodes and assigning channels: After determining the set of possible links, filter the set to obtain the minimum set which satisfies the total coverage of the region. Here we define a multi objective optimization approach in which we minimise the number of outdoor nodes, minimise the total number of nodes, when maximising the throughput in the respect of coverage requirement. To benefit from the multi-channel of nodes we can study how to apply the graph colouring approach [19] when using non-interfering channels (for Example 1, 6, and 11). Therefore, Mix-Rx-Tx interference that is due to simultaneous transmissions and receptions on the same node may be avoided.

Applying routing: Energy, end to end delay, and the throughput are of interest. We will evaluate three routing protocols: *Ad hoc* on Demand Distance Vector (AODV), Optimized Link State Routing (OLSR) and Hybrid Wireless Mesh Protocol (HWMP) according to these three constraints after selecting the suitable routing metrics and an energy consumption model. Eventually, we will try to optimize the best routing protocols. Evaluation will be made using Network Simulator 3 (NS3).

At the end of this step, verifications should be made in order to check if the QoS and coverage requirements are met.

Verifications: Verifications consist to check if the requirements in term of coverage and throughput are met. If it is not the case, distances are reduced to improve the throughput. Outdoor nodes are re-determined and Step 9 is performed again. This circle is done until the requirements are met. Therefore, the list of components is produced.

**Step 10: Cost Estimation**

After the planning, the list of component with dominant cost will be generated. Based on it and using distribution probabilities, a framework for the cost estimation will be built.

## 5. Conclusions

In rural Africa, wireless mesh network is very important for the development. The lack of methodology could result in very poor performance and lack of sustainability. This work is an approach for planning a suitable wireless mesh network for these regions. In this paper, we proposed an approach in ten steps based on indoor/outdoor deployment to reduce the overall cost of the network. These steps are: Analysing the region, identifying coverage requirement, identifying applications and services, estimating quality of service requirement, estimating elevation profile, identifying indoor candidate position, selecting network technology, determining outdoor candidate position, planning the network and estimation the overall cost.

Deploying such a network is very helpful for the local community. New opportunities could be provided both for the MCTs and clients. Clients do not need any more to move to the MCT, since they could access the network from their home. For the MCT, by this way, insufficiency of resource could be overcome and the incomes of the MCT can increase. Additionally, although there is a radio in some MCTs, this mean of communication is one-way direction (from MCT to inhabitant). To serve as a pillar for the development of the local community, the MCT should provide a platform for participatory approach, where inhabitants can get information from MCT and also provide information to MCT. This platform could help to share information about rural activities like agriculture, forestry, fisheries, and also to educate and train rural population.

The next relevant parts of this work are to describe and to perform Steps 7 to 10: Selecting network technology, identifying outdoor candidate positions, network planning, and cost estimation.